\newcounter{myctr}
\def\myitem{\refstepcounter{myctr}\bibfont\noindent\ifnum\themyctr>9\else\phantom{0}\fi\hangindent17pt\themyctr.\enskip}

\documentclass{ws-ijqi}
\usepackage{hyperref}
\usepackage[super,sort,compress]{cite}
\usepackage[numbers,sort&compress]{natbib}
\usepackage{url}
\begin{document}

\title{QUANTUM ENHANCED IMAGING OF NON-UNIFORM \\
REFRACTIVE PROFILES}

\author{GIUSEPPE ORTOLANO\footnote{giuseppe.ortolano@polito.it} $^{1,2}$, IVANO RUO BERCHERA$^1$ \vspace{10pt} }

\address{$^1$Quantum metrology and nano technologies division, \\ INRiM - Istituto Nazionale di Ricerca Metrologica, \\ Strada delle Cacce 91, 10135 Torino, Italy}
\address{$^2$DISAT - Dipartimento Scienza Applicata e Tecnologia, \\Politecnico di Torino, Corso Duca degli Abruzzi 24,\\
10129 Torino, Italy}
\author{ENRICO PREDAZZI}
\address{Dipartimento di Fisica - Università di Torino \\
                               and \\
           Sezione INFN di Torino - Via P. Giuria 1 - Torino - Italy}

\maketitle

\begin{abstract}
In this work quantum metrology techniques are applied to the imaging of objects with a non-uniform refractive spatial profile. A sensible improvement on the classical accuracy is shown to be found when the "Twin Beam State" (TWB) is used. In particular exploiting the multimode spatial correlation, naturally produced in the Parametric Down Conversion (PDC) process, allows a 2D reconstruction of complex spatial profiles, thus enabling an enhanced imaging. The idea is to use one of the spatially multimode beam to probe the sample and the other as a reference to reduce the noise. A similar model can be also used to describe wave front distortion measurements. The model is meant to be followed by a first experimental demonstration of such enhanced measurement scheme.
\end{abstract}


%\tableofcontents  % optional

\markboth{Giuseppe Ortolano, Ivano Ruo Berchera, Enrico Predazzi}
{Quantum enhanced imaging of non uniform refractive profiles}

\section{Introduction}	%) A SECTION HEADING
\begin{figure}[t]
\centerline{\includegraphics[width=\textwidth]{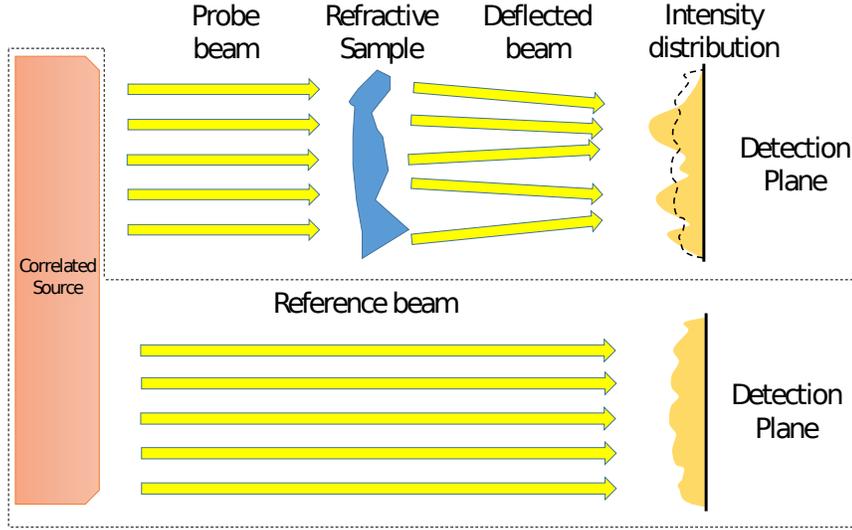}}
\vspace*{8pt}
\caption{\emph{Wide field imaging of a refractive profile}. A sample is illuminated by a spatial multimode beam. At each position the beam is deflected at a different angle, altering the intensity distribution. In the dashed shape the configuration in which the beam is correlated to another, used as reference, is pictured.}\label{wf}
\end{figure}
In recent years quantum states of light have been proven successful in the enhancement of a variety of measurement schemes \cite{Genovese_2016}, such as undetected photon imaging \cite{Lemos_2014}, quantum illumination\cite{Lopaeva_2013, Zhang_2015, Zhang_2019}, super resolution \cite{Gatto_2014,Classen_2017}, ghost imaging \cite{Pittman_1995, Puddu_2005, Meyers_2008, Bina_2013, Brida_2011, Meda_2015}, interferometry \cite{Aasi_2013,Berchera_2013,Pradyumna_2018,Schafermeier_2018} and absorption imaging \cite{Meda_2017,Berchera_2019,Knyazev_2019,Sabines-Chesterking_2019}. In particular a fundamental limit in the accuracy of classical schemes is the Shot Noise Limit (SNL) \cite{Giovannetti_2004, Giovannetti_2011}, that bounds the uncertainty in the estimation of a parameter to scale as the inverse square root of the photons involved. Schemes that enable to surpass the SNL are of paramount importance in settings where the energy that can be used is limited, as it is the case, for example, when dealing with biological samples \cite{Taylor_2016} that could be damaged by the radiation. Sub SNL measurements have been realized, using squeezed states of light, for interferometry \cite{Demkowicz_2014,Schnabel_2017}, beam displacement measurements \cite{Barnett_2003,Pooser_2015,Treps_2003}, and recently sub SNL wide field absorption imaging has been achieved \cite{Brida_2010,Samantaray_2017,Knyazev_2019}, using quantum correlated states. A state often used in such schemes is the Twin Beam (TWB) state produced by the process of Parametric Down Conversion (PDC) \cite{Burnham_1970,Jakeman_1986} or four wave mixing \citep{Glorieux_2011,Pooser_2016}. In PDC a laser pump interacts with a non-linear crystal creating, as a result, a pair of photons correlated both in position and momentum. This state is particularly interesting not only because the use of quantum correlations allows a reduction of the uncertainty of an estimation below the SNL, but also because of the spatial multimode nature of the PDC process, that automatically enables wide field imaging, meaning that a 2D spatial profile can be imaged with a single exposure. It can be expected that the TWB state, similarly as it is in the case of absorption imaging, can be used to achieve sub SNL measurements of non uniform refractive profiles and aim of this work is, in fact, to investigate the improvements that the use of quantum correlations would bring to such measurements.  Classically different techniques are used to image the refractive profile of an object. Between those the Schilieren scheme\cite{Settles_2006} focuses on the imaging of the gradient of the refractive profile $\nabla n$.  Considering a beam interacting with the object, using ray optics \citep{Marchand_1978}, can be seen that a deflection of a certain angle is produced, proportional to $\nabla n$ and in the Schilieren configuration this is in turn proportional to the difference in detected intensity, with and without the object. Thus, for each point of the object the angle of deflection is retrieved measuring the change in the intensity distribution at the detection plane, by means of a multipixel detector. \\
In this paper, similarly, we analyze the possible quantum advantage achievable in a scheme where the deflection is estimated by a measurement of the intensity distribution, so that the uncertainty of the estimation depends on the statistics of the detected photons. This problem is similar to the beam displacement problem analyzed in Ref.\cite{Pooser_2015}, where the entire beam is deflected of a certain angle and it is detected by a quadrant detector. The difference is that the structure causing the deflection in our case is more complex, in the sense that at each position of the sample incoming light is deflected at a different angle, or no angle at all, as pictured in figure \ref{wf}. The object is considered to be illuminated by a spatially incoherent source with a certain pattern, e.g. the TWB state. The results, after the interaction, is a measured intensity distribution where deflected and non deflected parts    of the probe pattern  sum up in intensity at each pixels. Interference effects are not considered here given the incoherent properties of the multimode source.
\begin{figure}[t]
\centerline{\includegraphics[width=\textwidth]{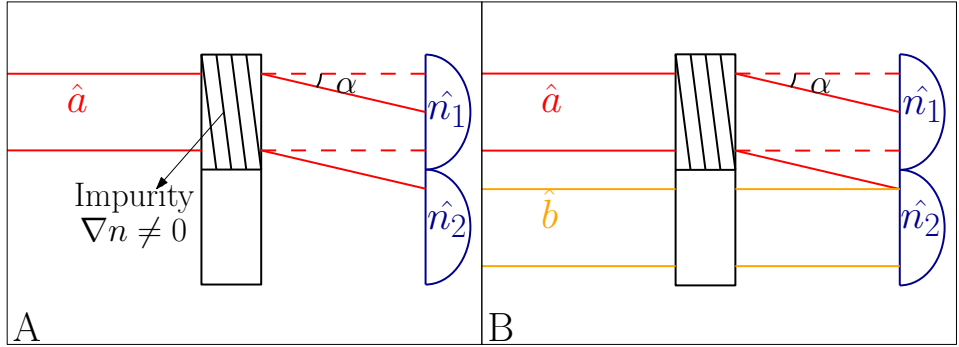}}
\vspace*{8pt}
\caption{\textbf{A}. \emph{Deflection of a single mode due to an impurity}. A single mode, labeled $\hat{a}$, interacts with a test object with a gradient in the refractive index $\nabla n$. As a result $\hat{a}$ is deflected of an angle $\alpha$. In turn this deflection will cause a shift in the detected position of the photons. \textbf{B}. \emph{Deflected mode with disturbance}. The difference with the previous scheme is the presence of a second mode, labeled as $\hat{b}$, and considered independent from $\hat{a}$. Due to the deflection some photons from $\hat{a}$ will be detected in the same position as photons from $\hat{b}$.}\label{cp}
\end{figure}
\section{The Model}
 The analysis of the interaction of the beam with the object can be carried out from a phenomenological point of view as depicted in figure \ref{cp}.A. \\
In the simplified scheme pictured a single mode, labeled $\hat{a}$ goes through a region with non-uniform refractive index, called an impurity, and, as a result, is deflected downwards of an angle $\alpha$. In turn at the detection plane, close to the object, photons will be detected in a shifted position. The detector are positioned such that the one labeled "1" intercepting the fist mode, when unperturbed, while an adjacent detector of the same size, labeled "2", receives photons only when the photon is deflected. The deflection is assumed small enough that the beam never exceeds the position of detector 2 at detection. In figure \ref{cp}.B a second mode, labeled $\hat{b}$ and considered independent from the first mode, is added, so that detector 2 in this case collects photons from $\hat{b}$ but also part of the photons from $\hat{a}$ due to the deflection. This last configuration mimics the situation one have in wide field imaging where the object can be illuminated simultaneously by different modes at different positions. The following analysis refers to this elementary scheme, but the situation can be generalized to the situation in which a gradient is present all across the object, producing local deflection.\\
We develop a quantum statistical model in which the deflection in figure \ref{cp}.B is represented as the result of a beam splitter (BS) acting on the mode $\hat{a}$ as showed in figure \ref{dbs}.A. The BS is characterized by its transmission coefficient $\tau$, the fraction of transmitted photons. The angle of deflection is then proportional to the reflectance $1-\tau$ where the constant of proportionality depends on the particular spatial distribution of the mode. Estimating  the angle of deflection of figure \ref{cp}.B is then equivalent to the estimation of the coefficient $\tau$ in the scheme \ref{dbs}.A.
\subsection{Direct scheme and SNL}\label{d}
\begin{figure}[t]
\centerline{\includegraphics[width=\textwidth]{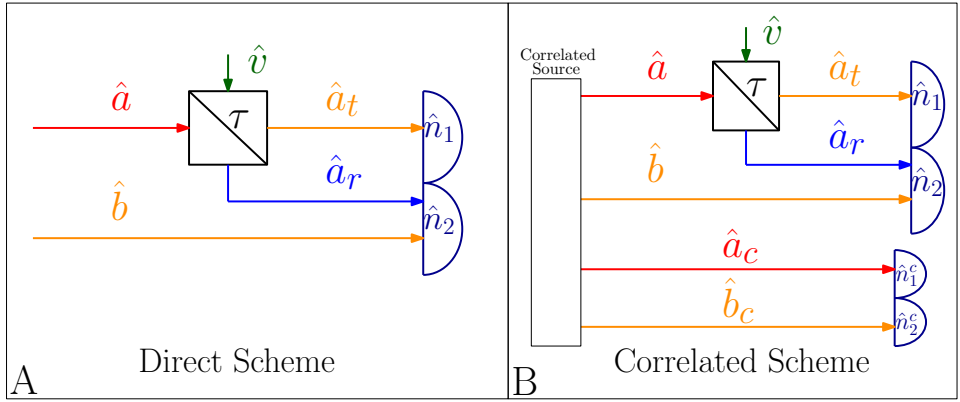}}
\vspace*{8pt}
\caption{\textbf{A}.\emph{Model of beam deflection}. Schematic representation of the situation of fig. \ref{cp}.B. The deflection of the beam is modeled with a BS of transmission $0\leq\tau\leq1$, where $1-\tau$ is proportional to the angle of deflection $\alpha$.  \textbf{B}. \emph{Correlated scheme}. The scheme pictures a deflection measurement. A correlate source is used to produce pairs of correlated modes, $\hat{a}$ correlated to $\hat{a}_c$ and $\hat{b}$ to $\hat{b}_c$. $\hat{a}$ and $\hat{b}$ probe the object, while their respective correlated modes are used as reference.}\label{dbs}
\end{figure}

Referring to the configuration of figure \ref{dbs}.A the estimation of $\tau$ can be carried out using the estimator $\hat{E}$:
\begin{equation}
\hat{E}=\frac{\hat{n}_1 -  \hat{n}_2 }{\hat{n}_1 +  \hat{n}_2 } \label{es}
\end{equation}
where $\hat{n}_1$ and $\hat{n}_2$ are the photon number operators detected from detectors 1 and 2 respectively. The choice of this estimator, where the role of the denominator is to attenuate the fluctuations, follows from the fact that it allows to reach the Ultimate Quantum Limit in the estimation of a BS parameter when the second mode $\hat{b}$ is not considered \citep{Monras_2007}. The estimator $\hat{E}$ is defined using a ratio of operator and is mean value can be found expanding equation  \ref{es}, for small fluctuations around the operator mean value, that at the zero-th order is just:
\begin{equation}
\langle\hat{E}\rangle=\Big\langle\frac{\hat{n}_1 -  \hat{n}_2 }{\hat{n}_1 +  \hat{n}_2 }\Big\rangle \approx\frac{\langle \hat{n}_1 \rangle - \langle \hat{n}_1 \rangle}{\langle \hat{n}_1 \rangle + \langle \hat{n}_2 \rangle}= \frac{(2\tau - 1) N_a - N_b}{N_a + N_b} \label{ev}
\end{equation}
where $N_a=\langle \hat{n}_a \rangle=\langle \hat{a}^\dagger\hat{a} \rangle$ and $N_b=\langle \hat{n}_b \rangle=\langle \hat{b}^\dagger\hat{b} \rangle$ are the mean number of photons in modes $\hat{a}$ and $\hat{b}$.
 All the mean values $\langle \cdot \rangle$ are taken on the initial state of the field, $\rho_{a}\otimes\rho_{b}$. Those states will be specified by means of their photons statistics, and given the  physical configuration under analysis, from now on we will consider $\rho_{a}$ and $\rho_{b}$ equal, as the two modes are produced by the same source. An estimation of $\tau$ can be found solving equation \ref{ev}, using the fact that $N_a$ and $N_b$ can be considered parameters as they can be determined with arbitrary accuracy in a preliminary characterization of the experimental apparatus, in absence of the sample under test. \\
The variance $\langle \Delta^2 \hat{E}\rangle$ can be obtained with the propagation of the uncertainty on $\hat{n}_1$ and $\hat{n}_2$ and  expressed in terms of the statistic of the input modes $\hat{a}$ and $\hat{b}$. \\
From the well known BS relations $\hat{a}_t=\sqrt{\tau}\hat{a}+i\sqrt{(1-\tau)}\hat{v}$ and $\hat{a}_r=i\sqrt{(1-\tau)}\hat{a}+\sqrt{\tau}\hat{v}$, the statistic of the transmitted and reflected modes $\hat{a}_t$ and $\hat{a}_r$ is easily found to be:
\begin{align}
\langle \hat{n}_t \rangle  &= \langle \hat{a}_t^\dagger\hat{a}_t \rangle  =\tau N_a  &\langle \hat{n}_r \rangle &= \langle \hat{a}_r^\dagger\hat{a}_r \rangle  = (1-\tau)N_a  \nonumber \\
\langle \Delta^2 \hat{n}_t \rangle &= \tau N_a( \tau F +1-\tau )&\langle \Delta^2\hat{n }_r \rangle &= N_a(1-\tau)(F(1-\tau)+\tau) \label{bs} \\	
\langle \Delta\hat{n}_t\Delta\hat{n}_r \rangle  &= \tau(1-\tau)N_0(F-1) \nonumber
\end{align}
The Fano factor \cite{Fano_1947} $F=\langle \Delta^2\hat{n }_a \rangle/\langle \hat{n }_a \rangle$ was introduced to characterize the statistic of the input state. States with $F<1$, i.e. characterized by sub-Poissonian fluctuation, are considered non-classical states of light \cite{Mandel_1995}. The statistic of $\hat{n}_1$ follows directly from relations \ref{bs} since from scheme \ref{dbs}.A it coincides with $\hat{n}_t$. To determine the statistic of $\hat{n}_2$ we use the fact that $\hat{a}$ and $\hat{b}$ are independent so that we have:
\begin{align}
\langle \hat{n}_2 \rangle &= \langle \hat{n}_b \rangle + \langle \hat{n}_r \rangle \nonumber \\
\langle \Delta\hat{n}_1\Delta\hat{n}_2 \rangle &= \langle \Delta\hat{n}_t\Delta\hat{n}_r \rangle \\
\langle \Delta^2\hat{n}_2 \rangle &= \langle \Delta^2\hat{n}_b \rangle + \langle \Delta^2\hat{n}_r \rangle \nonumber
\end{align}
So that for $\hat{n}_2$ we get:
\begin{align}
\langle \hat{n}_2 \rangle &= N_b + (1-\tau)N_a \nonumber  \\
\langle \Delta^2\hat{n}_2 \rangle &= F N_b + (1-\tau)^2 F N_a +\tau(1-\tau)N_a \label{v2} \\	
\langle \Delta\hat{n}_1\Delta\hat{n}_2 \rangle &=\tau(1-\tau)\big(F N_a  - N_a \big) \nonumber
\end{align}
Using equations \ref{bs} and \ref{v2} we can propagate the uncertainty from equation \ref{es}. Assuming $N_a=N_b=N$ we have:
\begin{equation}
\langle\Delta^2\hat{E}\rangle \approx \frac{F \tau^2 }{2N} + \frac{\tau(1-\tau)}{N}
\end{equation}
This uncertainty can be propagated to the parameter $\tau$ as:
\begin{equation}
\Delta \tau= \frac{\sqrt{\langle\Delta^2 \hat{E} \rangle}}{|\partial \langle \hat{E}\rangle /\partial \tau|} \label{pr}
\end{equation}
So that:
\begin{equation}
\Delta \tau  = \sqrt{\frac{F \tau^2  }{2 N}  + \frac{\tau(1-\tau)}{N}} \label{vt}
\end{equation}
The minimum fluctuation that can be achieved with "classical" states is the one obtained with coherent states, with $F=1$, setting the SNL for this scheme:
\begin{equation}
\Delta \tau_{SNL}  = \sqrt{ \frac{ \tau^2 }{2 N} + \frac{\tau(1-\tau)}{N}} \label{vt}
\end{equation}
\subsection{Correlated scheme}
In order to take advantage of quantum correlations, we propose another scheme, depicted in figure \ref{dbs}.B. A source is used to produce spatially separated pairs of correlated modes. In picture \ref{dbs}.B the modes testing the object, $\hat{a}$ and $\hat{b}$, are correlated to the modes $\hat{a}_c$ and $\hat{b}_c$ respectively, that act as a reference.  The aim of this scheme is to exploit correlations in photon numbers to improve the accuracy over the direct scheme. The degree of correlation, for a pair of generic modes $\hat{i}$ and $\hat{j}$, is expressed by the noise reduction factor \citep{Brida_2010} $\sigma$ defined as:
\begin{equation}
	\sigma=\frac{\langle \Delta^2 (\hat{n}_i-\hat{n}_j) \rangle}{\langle \hat{n_i}+\hat{n_j} \rangle} \label{nr}
\end{equation}
With this configuration the parameter $\tau$ can be computed using the estimator $\hat{E}_{C}$:
\begin{equation}
\hat{E}_{C}=\frac{\hat{n}_1 -  (\hat{n}_2 - \hat{n}_2^c) }{\hat{n}_1^c} \label{edf}
\end{equation}
The choice of this estimator is arbitrary but motivated by the fact that the correlation of $\hat{n}_2$ and $\hat{n}_2^c$ should allow to reduce the fluctuation of the bracket term at the numerator, meanwhile normalizing by $\hat{n}_1^c$ compensates for the fluctuation of $\hat{n}_1$. \\
For small fluctuation in photon numbers, the mean value can be approximated, as done before as:
\begin{equation}
\langle\hat{E}_{C} \rangle \approx \frac{\langle \hat{n}_1 - (\hat{n}_2 - \hat{n}_2^c)\rangle }{\langle\hat{n}_1^c\rangle} = 2\tau-1
\end{equation}
The calculation of the uncertainty is similar to the one showed in the previous section and will not be reported. The result is:
\begin{equation}
\Delta \tau_C= \sqrt{\frac{\tau(1-\tau)}{N}+\frac{(2\tau-1)^2\sigma}{4N} + \frac{\sigma}{2N } } \label{vc}
\end{equation}
that depends only on the measured mean number of photons $N$ in the reference beam and on the measured noise reduction factor in  absence of the sample's perturbation.
\section{Results and Discussion}
From equations \ref{vt} and \ref{vc} a  comparison of the performance  in the estimation with different input states can be made.  In particular for the direct scheme, of section \ref{d} we consider each mode of the multimode beam to be , alternatively, in one of the following states::
\begin{itemize}
\item The \emph{Fock state}, eigenstate of the photon number operator of the field so that $F_{Fock}=0$
\item The \emph{coherent state}, eigenstate of the annihilation operator with a Poissonian photon number distribution, hence $F_{coh}=1$
\item The \emph{thermal state}, a mixed state characterized by the Bose-Einstein distribution at thermal equilibrium, $P(n) = \frac{N^n}{(1+N)^n}$ \citep{Scully_1997}, having then $F_{th}=1+N$, $N$ being the main number of photons.
\end{itemize}
The differential scheme will analyzed in the case of the \emph{TWB state}:
\begin{equation}
|\psi\rangle_{\text{TWB}}=\sum_{n} c(n)|n \rangle_{\vec{k}_t,\omega} |n \rangle_{-\vec{k}_t,-\omega} \label{tw}
\end{equation}
where $\vec{k}_t$ and $\omega$ are the transverse momentum and frequency of the mode and $|c(n)|^2$ is a thermal like distribution with parameter $N$. From \ref{tw} it is clear that the quantum nature of the state resides in its entanglement, as tracing out either one of the modes would give a thermal statistic for the other. Moreover it is easy to see that for this state, due to the perfect photon number correlation, the noise reduction factor is $\sigma_{\text{TWB}}=0$.  \\
\begin{figure}[t]
\centerline{\includegraphics[width=\textwidth]{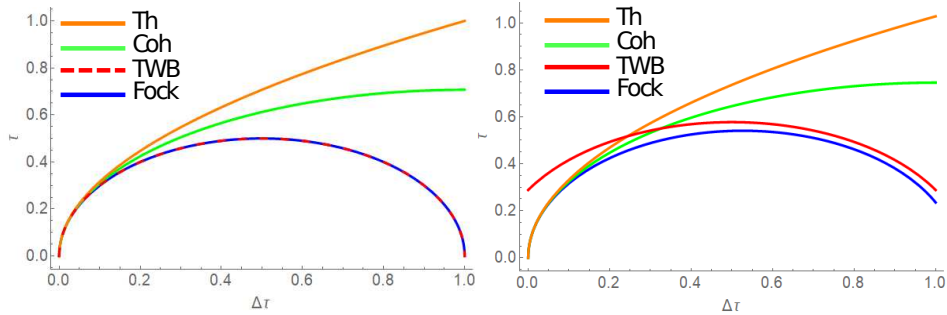}}
\vspace*{8pt}
\caption{\textbf{A}.\emph{Uncertainty on the estimation of the beam splitter parameter $0<\tau<1$, modelling a beam deflection}. Referring to scheme \ref{dbs}.A the input states are Fock(blue), coherent(green) and thermal state(orange). The TWB state result, the dashed line in red coinciding with the Fock state, refers to scheme \ref{dbs}.B. \textbf{B}.\emph{Uncertainty on the estimation of the beam splitter parameter $0<\tau<1$, modelling a beam deflection, with efficiency $\eta=0.9$}. The uncertainty of the measurement scheme \ref{dbs}.A is plotted in the case of optical efficiency $\eta=0.9$, meaning that a fraction $1-\eta$ of the initial number of photons are lost. The input states considered are Fock(blue), coherent(green) and thermal(orange) state. The TWB state result, plotted in red, refers to scheme \ref{dbs}.B, where the efficiency is considered $\eta=0.9$ in both the probe and reference channel.}\label{nl}
\end{figure}
In figure \ref{nl}.A the uncertainty $\Delta \tau$ on the estimation is plotted against the parameter $\tau$ in the case of each of the states discussed. The curves are obtained by simply substituting the Fano factor of the different states considered in equation \ref{vt}, for the direct scheme, and  $\sigma=0$ in \ref{vc} for the correlated case. The minimum uncertainty attainable in the estimation of a BS parameter \cite{Monras_2007} is reached, for every value of $\tau$ by both the TWB and the Fock state. It is not surprising that the Fock state reaches the lower bound to the uncertainty, since the estimation  is based on photon number measurement, for which this state has no noise. When lossless channels are considered, the use of quantum correlations allows to erase the quantum noise present in the probe beam, by exploiting the the information on the photon number fluctuation measured in the reference beam, reproducing the situation in which the field is prepared in a Fock state. The coherent state, plotted in green, is a useful reference for the performance of the TWB state, since as mentioned before, the former represent the SNL and so the limit achievable with classical states. The advantage of TWB over the SNL gets more evident in the region of high $\tau$, corresponding to low deflections. The thermal state is, as expected, the worst one and is reported to show the disadvantage in the use of of light modes in noisier states unless quantum correlation are used. \\
Up until now, possible photon losses have not been considered,  although they are unavoidable in any real optical scheme.  Since optical losses are random processes, that add a certain amount of noise, sub-Poissonian behavior and quantum correlations are strongly affected by them.  The Fano factor and the NRF measured in case of a fraction $0\leq1-\eta\leq1$ of photons lost in the channel are:
\begin{align}
&F_{\eta}=\eta F+1 - \eta &\sigma_{\eta}=\eta \sigma +1 -\eta \label{ei}
\end{align}
where for $\sigma_{\eta}$ equal losses on the correlated channels have been assumed. \\
In figure \ref{nl}.B the uncertainty is reported in the case of an high, but not perfect, efficiency, $\eta=0.9$, evaluated by substituting expressions \ref{ei} into equations \ref{vt} and \ref{vc}. In this scenario the performance of the TWB state does not coincide anymore with the one of the Fock state but  it becomes slightly worse. An interesting feature is that the uncertainty of the TWB estimation does not approach zero as $\tau \rightarrow 0$ and as a consequence the TWB performs worst than any other configuration in the high deflections region. This is a consequence of the choice of \ref{edf} as an estimator, and can be eliminated with a different one. The advantage, however, of \ref{edf} over other tested estimators, and the reason why it has been chosen here, is that it allows to improve the sensitivity for small deflections, the one we are more interested in. In this region the TWB state approaches the result of the Fock state, even in presence of losses, and gives a sensible improvement over the SNL. \\
\begin{figure}[t]
\centerline{\includegraphics[width=0.7\textwidth]{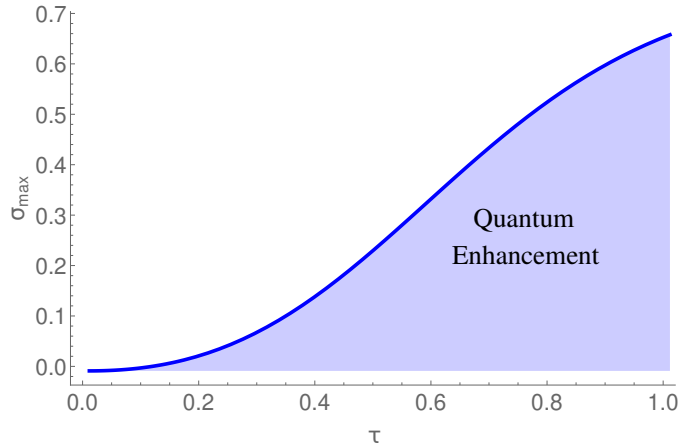}}
\vspace*{8pt}
\caption{\emph{NRF needed to beat the SNL}. The plot represent the maximum value of noise reduction factor $\sigma$ that a correlated state, used in scheme \ref{dbs}.B, can have in order to have an advantage over the SNL \ref{vt}.}\label{nra}
\end{figure}
Finally in figure \ref{nra} we report the maximum value of the detected noise reduction factor $\sigma_{max}$, required to have an advantage over the SNL, i.e. $ \Delta \tau_C(\sigma_{max})< \Delta \tau_{SNL} $. It is interesting to notice how, the higher the deflection (smaller values of $\tau$), the stronger the correlation has to be to grant an advantage over the classical limit. Moreover even in the limiting case of no deflection, $\tau=1$, a NRF $\approx0.7$ is still required, a value well below the limit achievable by classical correlations $\sigma_{class}\geq1$.
\section{Conclusions}
In this work, a simple quantum model describing the measurement of a refractive profile, based on the change of the intensity distribution of a beam after the interaction with a sample, has been elaborated to investigate a possible quantum enhancement in the sensitivity. The deflections caused on the spatially multimode beam interacting with the test object, were modeled using a beam splitter transformation with transmission coefficient $\tau$, where the angle of deflection $\alpha$ is proportional to $1-\tau$.  A direct measurement scheme was compared to a correlated one, where quantum correlations are used to improve the accuracy. In particular we found that the TWB state, a state characterized by entanglement in photon number between pairs of spatio-temporal modes, overcomes  the Shot noise limit (SNL) both in the ideal lossless case, reported in figure \ref{nl}.A and in presence of losses shown in figure \ref{nl}.B. Moreover, we have shown that only a correlation level well above the classical bound (noise reduction factor $\sigma_{max}<0.7$) allows to overcome the SNL, as reported in figure \ref{nra}. \\
This results show the possibility to reach a quantum enhancement for wide field imaging of refractive profiles inducing an intensity perturbation in the near field, using a TWB configuration. The analysis performed in this work is meant to be followed by a wide field experimental realization of the differential scheme with the TWB state. Twin beams are, in fact, currently routinely generated in quantum optics laboratories, and they have already been used for sub shot noise imaging of absorption profiles. Thus the scheme suggested in this work for refractive profile measurements is feasible with the current technology. \\
Realizing sub SNL wide field imaging is especially important when there is a limit on the energy that can be used to probe samples. For this reason sub SNL imaging of refractive profiles would have useful application, for example, in the analysis of quasi transparent biological sample, giving complementary information to the one obtained using other measurements.
\section*{Acknowledgments}
This work was funded through the EMPIR project 17FUN01-BeCOMe (The EMPIR initiative is funded by the European Union Horizon 2020 research and innovation programme and co-financed  by the EMPIR participating States) and through Horizon 2020 research and innovation programme  under grant agreement number 862644 (FET-open- QUARTET).
\bibliographystyle{ws-ijqi}
\bibliography{bib}{}

\begin{thebibliography}{10}

\bibitem{Genovese_2016}
M.~Genovese, {\em Journal of Optics} {\bf 18} (jun 2016) p. 073002.

\bibitem{Lemos_2014}
G.~B. Lemos, V.~Borish, G.~D. Cole, S.~Ramelow, R.~Lapkiewicz and A.~Zeilinger,
  {\em Nature} {\bf 512} (Aug 2014) 409 EP .

\bibitem{Lopaeva_2013}
E.~D. Lopaeva, I.~Ruo~Berchera, I.~P. Degiovanni, S.~Olivares, G.~Brida and
  M.~Genovese, {\em Phys. Rev. Lett.} {\bf 110} (Apr 2013) p. 153603.

\bibitem{Zhang_2015}
Z.~Zhang, S.~Mouradian, F.~N.~C. Wong and J.~H. Shapiro, {\em Phys. Rev. Lett.}
  {\bf 114} (Mar 2015) p. 110506.

\bibitem{Zhang_2019}
Y.~Zhang, D.~England, A.~Nomerotski, P.~Svihra, S.~Ferrante, P.~Hockett and
  B.~Sussman, Multidimensional quantum illumination via direct measurement of
  spectro-temporal correlations  (2019).

\bibitem{Gatto_2014}
D.~Gatto~Monticone, K.~Katamadze, P.~Traina, E.~Moreva, J.~Forneris,
  I.~Ruo-Berchera, P.~Olivero, I.~P. Degiovanni, G.~Brida and M.~Genovese, {\em
  Phys. Rev. Lett.} {\bf 113} (Sep 2014) p. 143602.

\bibitem{Classen_2017}
A.~Classen, J.~von Zanthier, M.~O. Scully and G.~S. Agarwal, {\em Optica} {\bf
  4} (Jun 2017) 580.

\bibitem{Pittman_1995}
T.~B. Pittman, Y.~H. Shih, D.~V. Strekalov and A.~V. Sergienko, {\em Phys. Rev.
  A} {\bf 52} (Nov 1995) R3429.

\bibitem{Puddu_2005}
E.~Puddu, A.~Allevi, A.~Andreoni and M.~Bondani, {\em Opt. Lett.} {\bf 30} (Jun
  2005) 1294.

\bibitem{Meyers_2008}
R.~Meyers, K.~S. Deacon and Y.~Shih, {\em Phys. Rev. A} {\bf 77} (Apr 2008) p.
  041801.

\bibitem{Bina_2013}
M.~Bina, D.~Magatti, M.~Molteni, A.~Gatti, L.~A. Lugiato and F.~Ferri, {\em
  Phys. Rev. Lett.} {\bf 110} (Feb 2013) p. 083901.

\bibitem{Brida_2011}
G.~Brida, M.~V. Chekhova, G.~A. Fornaro, M.~Genovese, E.~D. Lopaeva and I.~R.
  Berchera, {\em Phys. Rev. A} {\bf 83} (Jun 2011) p. 063807.

\bibitem{Meda_2015}
A.~Meda, A.~Caprile, A.~Avella, I.~Ruo~Berchera, I.~P. Degiovanni, A.~Magni and
  M.~Genovese, {\em Applied Physics Letters} {\bf 106}  (2015) p. 262405.

\bibitem{Aasi_2013}
J.~Aasi {\em et~al.}, {\em Nature Photonics} {\bf 7} (Jul 2013) 613 EP .

\bibitem{Berchera_2013}
I.~Ruo~Berchera, I.~P. Degiovanni, S.~Olivares and M.~Genovese, {\em Phys. Rev.
  Lett.} {\bf 110} (May 2013) p. 213601.

\bibitem{Pradyumna_2018}
S.~T. Pradyumna, E.~Losero, I.~Ruo-Berchera, P.~Traina, M.~Zucco, C.~S.
  Jacobsen, U.~L. Andersen, I.~P. Degiovanni, M.~Genovese and T.~Gehring,
  Quantum-enhanced correlated interferometry for fundamental physics tests
  (2018).

\bibitem{Schafermeier_2018}
C.~Sch\"{a}fermeier, M.~Je\v{z}ek, L.~S. Madsen, T.~Gehring and U.~L. Andersen,
  {\em Optica} {\bf 5} (Jan 2018) 60.

\bibitem{Meda_2017}
A.~Meda, E.~Losero, N.~Samantaray, F.~Scafirimuto, S.~Pradyumna, A.~Avella,
  I.~Ruo-Berchera and M.~Genovese, {\em Journal of Optics} {\bf 19} (aug 2017)
  p. 094002.

\bibitem{Berchera_2019}
I.~R. Berchera and I.~P. Degiovanni, {\em Metrologia} {\bf 56} (jan 2019) p.
  024001.

\bibitem{Knyazev_2019}
E.~Knyazev, F.~Y. Khalili and M.~V. Chekhova, {\em Opt. Express} {\bf 27} (Mar
  2019) 7868.

\bibitem{Sabines-Chesterking_2019}
J.~Sabines-Chesterking, A.~R. McMillan, P.~A. Moreau, S.~K. Joshi, S.~Knauer,
  E.~Johnston, J.~G. Rarity and J.~C.~F. Matthews, {\em Opt. Express} {\bf 27}
  (Oct 2019) 30810.

\bibitem{Giovannetti_2004}
V.~Giovannetti, S.~Lloyd and L.~Maccone, {\em Science} {\bf 306}  (2004) 1330.

\bibitem{Giovannetti_2011}
V.~Giovannetti, S.~Lloyd and L.~Maccone, {\em Nature Photonics} {\bf 5} (Mar
  2011) 222 EP .
\newblock Review Article.

\bibitem{Taylor_2016}
M.~A. Taylor and W.~P. Bowen, {\em Physics Reports} {\bf 615}  (2016) 1 .
\newblock Quantum metrology and its application in biology.

\bibitem{Demkowicz_2014}
R.~{Demkowicz-Dobrzanski}, M.~{Jarzyna} and J.~{Kolodynski}, {\em Progess in
  Optics} {\bf 60}  (2015) p. 345.

\bibitem{Schnabel_2017}
R.~Schnabel, {\em Physics Reports} {\bf 684}  (2017) 1 .
\newblock Squeezed states of light and their applications in laser
  interferometers.

\bibitem{Barnett_2003}
S.~Barnett, C.~Fabre and A.~Ma{\i}tre, {\em The European Physical Journal D -
  Atomic, Molecular, Optical and Plasma Physics} {\bf 22} (Mar 2003) 513.

\bibitem{Pooser_2015}
R.~C. Pooser and B.~Lawrie, {\em Optica} {\bf 2} (May 2015) 393.

\bibitem{Treps_2003}
N.~Treps, N.~Grosse, W.~P. Bowen, C.~Fabre, H.-A. Bachor and P.~K. Lam, {\em
  Science} {\bf 301}  (2003) 940.

\bibitem{Brida_2010}
G.~{Brida}, M.~{Genovese} and I.~{Ruo Berchera}, {\em Nature Photonics} {\bf 4}
  (April 2010) 227.

\bibitem{Samantaray_2017}
N.~Samantaray, I.~Ruo-Berchera, A.~Meda and M.~Genovese, {\em Light: Science
  \&Amp; Applications} {\bf 6} (Jul 2017) e17005 EP .
\newblock Original Article.

\bibitem{Burnham_1970}
D.~C. Burnham and D.~L. Weinberg, {\em Phys. Rev. Lett.} {\bf 25} (Jul 1970)
  84.

\bibitem{Jakeman_1986}
E.~Jakeman and J.~Rarity, {\em Optics Communications} {\bf 59}  (1986) 219 .

\bibitem{Glorieux_2011}
Q.~Glorieux, L.~Guidoni, S.~Guibal, J.-P. Likforman and T.~Coudreau, {\em Phys.
  Rev. A} {\bf 84} (Nov 2011) p. 053826.

\bibitem{Pooser_2016}
R.~C. Pooser and B.~Lawrie, {\em ACS Photonics} {\bf 3}  (2016) 8.

\bibitem{Settles_2006}
G.~Settles, {\em Schlieren \& Shadowgraph Techniques} (Springer, jul 2006).

\bibitem{Marchand_1978}
E.~Marchand, {\em Gradient index optics} (Academic Press, 1978).

\bibitem{Monras_2007}
A.~Monras and M.~G.~A. Paris, {\em Phys. Rev. Lett.} {\bf 98} (Apr 2007) p.
  160401.

\bibitem{Fano_1947}
U.~Fano, {\em Phys. Rev.} {\bf 72} (Jul 1947) 26.

\bibitem{Mandel_1995}
L.~Mandel and E.~Wolf, {\em Optical Coherence and Quantum Optics} (Cambridge
  University Press, 1995).

\bibitem{Scully_1997}
M.~O. Scully, {\em Quantum Optics} (Cambridge University Press, sep 1997).

\end{thebibliography}


\begin{thebibliography}{1}
\providecommand{\natexlab}[1]{#1}
\providecommand{\url}[1]{\texttt{#1}}
\expandafter\ifx\csname urlstyle\endcsname\relax
  \providecommand{\doi}[1]{doi: #1}\else
  \providecommand{\doi}{doi: \begingroup \urlstyle{rm}\Url}\fi

\bibitem[Meda et~al.(2017)Meda, Losero, Samantaray, Scafirimuto, Pradyumna,
  Avella, Ruo-Berchera, and Genovese]{Meda_2017}
A~Meda, E~Losero, N~Samantaray, F~Scafirimuto, S~Pradyumna, A~Avella,
  I~Ruo-Berchera, and M~Genovese.
\newblock Photon-number correlation for quantum enhanced imaging and sensing.
\newblock \emph{Journal of Optics}, 19\penalty0 (9):\penalty0 094002, aug 2017.
\newblock \doi{10.1088/2040-8986/aa7b27}.
\newblock URL \url{https://doi.org/10.1088%2F2040-8986%2Faa7b27}.

\end{thebibliography}

\end{document}